\documentclass[showpacs, showkeys, 12pt, final, lengthcheck, nobibnotes, nofootinbib, longbibliography, onecolumn, aps, reprint]{revtex4-1}

\usepackage{amssymb,amsmath,xcolor,graphicx,xspace,colortbl,rotating}  
\usepackage{graphics}   

\DeclareGraphicsExtensions{.pdf,.eps,.ps,.png,.jpg,.jpeg}

\begin{document}
\title{Fast Entropy Estimation for Natural Sequences
}
\author{Andrew D. Back
}
\email[Contact email: ]{a.back@uq.edu.au}
\affiliation{School of ITEE, The University of Queensland, Brisbane, QLD, 4072 Australia.
}
\author{Daniel Angus
}
\affiliation{School of ITEE, The University of Queensland, Brisbane, QLD, 4072 Australia.
}
\author{Janet Wiles
}
\affiliation{School of ITEE, The University of Queensland, Brisbane, QLD, 4072 Australia.
}
\begin{abstract} It is well known that to estimate the Shannon entropy for
symbolic sequences accurately requires a large number of samples. When some aspects of the
data are known it is plausible       to attempt to use this to more efficiently compute entropy. A
number of methods       having various assumptions have been proposed which can be used to calculate
entropy for small sample sizes. In this paper, we examine this problem and
propose a method for estimating the Shannon entropy for a set of ranked symbolic ``natural''
events. Using a modified Zipf-Mandelbrot-Li law and a
new rank-based coincidence counting method, we propose an efficient
algorithm which enables the entropy to be estimated with surprising
accuracy using only a small number of samples. The algorithm is tested
on some natural sequences and shown to yield accurate results with very small amounts of
data.
\end{abstract}
\maketitle

\section{Introduction
}
 Machine learning methods typically rely on forming models based on
statistical properties of observed data. An area of importance in this regard is information
theoretic methods which involve computing Shannon entropy and mutual information. The idea that the
randomness of a message can give a measure of the information it
conveys formed the basis of Shannon's entropy theory which gives a means of assigning a value to the
information
carried within a message \cite{Shannon48},\cite{Shannon483}. The way in which Shannon formulated this principle is that,
given a single random variable
$x$
which may take
$M$
distinct values, and is in this sense symbolic, where each value occurs independently with
probability
$p\left (x_{i}\right ) ,$
$i \in [1 ,M] ,$
then the single symbol Shannon entropy is defined as:
\begin{equation}H_{1}(X) = -\sum \limits _{i =1}^{M}p(x_{i})\log _{2}(p\left (x_{i}\right ))
\end{equation}

This extends to the case where the probabilities of multiple symbols
occurring together are taken into account. The general \textit{N}-gram entropy, which is a
measure of the information due to the statistical probability of
$N$
adjacent symbols occuring consecutively, can be derived as
\begin{equation}H_{N}(X\vert B) = -\sum \limits _{i ,j}^{_{}}p(b_{i} ,x_{j})\log _{2}(p(x_{j}\vert b_{i}))
\end{equation}
where
$b_{i} \in \sum ^{N -1}$
is a block of
$N -1$
symbols,
$x_{j}$
is an arbitrary symbol following
$b_{i} ,$
$p(b_{i} ,x_{j})$
is the probability of the \textit{N}-gram
$(b_{i} ,x_{j}) ,$
$p(x_{j}\vert b_{i})$
is the conditional probability of
$x_{j}$
occurring after
$b_{i}$
and is given by
$p(b_{i} ,x_{j})/p(b_{i}).$

One of the limitations of computing entropy accurately is the dependence on large
amounts of data, even more so when computing \textit{N}-gram entropy. Estimates of
entropy based on letter, word and \textit{N}-gram
statistics have often relied on large data
sets \cite{Ebeling94}, \cite{Moreno2015}. The reliance on long data sequences to estimate the probability
distributions used to calculate entropy and attempts to
overcome this in coding schemes is discussed in \cite{schurmann-grassberger-96} where they provide an estimate of letter entropy
extrapolated for infinite text lengths. A method of estimating the number of samples required to
compute entropy was proposed in \cite{BackAngusWilesEstEntropy2018}
which showed that a very large number of samples may be required to do this accurately.

Various approaches to estimating entropy over finite sample sizes have
been considered. A method of computing the entropy of dynamical systems which corrects for
statistical fluctuations of the sample data over finite sample sizes has been proposed in \cite{Grassberger1988}. Estimation
techniques using small datasets have been
proposed in \cite{Bonachela2008}, and an online approach for
estimating entropy in limited resource environments was proposed
in \cite{Paavola2011}. Entropy estimation over
short symbolic sequences was considered in the context of dynamical time series models based on
logistic maps and correlated Markov chains, where an effective shortened sequence length was
proposed which accounted for the correlation effect \cite{lesne2009}. A novel approach for calculating entropy using the idea
of estimating probabilities from a quadratic function of the
inverse number of symbol coincidences was proposed in \cite{Montalvao2012}. A limitation of this method was that it assumed
equiprobable symbols. The difficulty of estimating entropy due to the heavy tailed distribution of
natural sequences has been recognized, where it has been shown that the bias using classical
estimators depends the sample size and the characteristics of the heavy-tailed distribution
\cite{Gerlach2016}. A Bayesian model approach to inferring the
probability distributions has been considered at length
in \cite{NemenmaNips2002} and
\cite{Schurmann2015}. A computationally
efficient method for calculating entropy based on a James-Stein-type shrinkage estimator
was proposed by Hausser and Strimmer in \cite{hausser2009entropy}.

In this paper, by considering a model for the probability distributions of natural
sequence data, we propose an extension to the algorithm in  \cite{Montalvao2012} which enables a fast method of estimating entropy using a
small number of samples. The proposed algorithm is derived in the subsequent sections and
simulations are given showing its effectiveness.

\section{Proposed Algorithm for Estimating Entropy
}

\subsection{Coincidence Counting For Equiprobable Symbols
}
To compute Shannon entropy by estimating the symbol probabilities using
conventional histogram plug-in methods is effective for small alphabet sizes, however for non
equiprobable symbols with a large alphabet size, a very large number of symbols may be required. For
a given alphabet size M, to estimate the entropy with some degree of accuracy it is normally
required to estimate the probabilities of M symbols. Another approach is to adopt a parametric model
of the symbol probabilities. In this case, the idea is to form an invertible model
$J(M)$
of the relationship between the model parameters and some observable statistical
feature of the data. Then, the model is inverted and the statistical features of the actual data are
observed which enables the model parameters and hence entropy to be estimated.

The method of coincidence detection is based on the idea that a
discrete (or symbolic) random variable
$x$
which may take on a finite number
$M$
of distinct values
$x_{i} \in \left \{x_{1} ,\ldots  ,x_{M}\right \}$
with probabilities
$p\left (x_{i}\right ) ,i \in [1 ,M] .$
Consider the case where
$p\left (x_{i}\right ) =p\left (x_{j}\right ) \forall i ,j \in [1 ,M] ,$
that is, the symbols are equiprobable. Hence we may proceed as follows. The probability
of drawing any symbol on the first try followed by any other different symbol on the second try,
that  is, any two non repeating symbols is
\begin{equation}\widetilde{F}(2;M) =\frac{M(M -1)}{M^{2}}
\end{equation}
and hence the probability of drawing any two repeating or identical symbols out of the
entire set is
\begin{equation}F(2;M) =1 -\frac{M(M -1)}{M^{2}}
\end{equation}
Extending this to
$n$
draws, the probability of drawing any symbol on the first try followed by any other
different symbol\protect\footnote{That is, the probability of no repeating symbols in the entire sequence.
The reason for this formulation, is that by excluding all repeating symbols, it enables us to
compute the probability of any repeating symbols over a given sequence and hence the exact
probability of a coincident event at a specific sample instance, which by definition in (\ref{feqn}), must be at the
$n$
th sample since we have discounted the probabilities up to the
$(n -1)$
th
sample.
} up to the \textit{$n$
th }draw up to
$n$
symbols is
\begin{equation}\widetilde{F}(n;M) =\frac{M(M -1) \cdots (M -n +1)}{M^{n}}
\end{equation}
Therefore, it follows that the probability of drawing any
$q_{n} \in [2 ,\ldots  ,n]$
identical symbols (ie one or more repeating symbols in any position) out of the entire
set is given by
\begin{equation}F(n;M) =1 -\frac{M(M -1) \cdots (M -n +1)}{M^{n}} \label{Fneqn}
\end{equation}
To compute the probability of a first coincidence occurring exactly at the
$n$
th symbol for
$1 <n <M ,$
means that it is necessary to compute the probability of drawing no repeating symbols
in the entire sequence up to the
$(n -1)$
th draw given by
$\widetilde{F}(n -1;M)$
and consequently drawing any
$q_{n -1} \in [2 ,\ldots  ,n -1]$
identical symbols              is given by
$F(n -1;M) .$
Hence the required probability is given by (\cite{Montalvao2012}):
\begin{equation}f(n;M) =F(n;M) -F(n -1;M) \label{feqn}
\end{equation}
The expectation of the discrete parameter
$n$
and its associated probability
$f(n;M)$
is given by:
\begin{align}E[n] &  =J(n;M) \label{Jeqn} \\
 &  =\sum \limits _{n =0}^{M}nf(n;M) \label{Jeqn2}\end{align}
Since
$n$
is a function of
$M ,$
define
\begin{equation}D(M) =E[n] .
\end{equation}
The innovative approach by \cite{Montalvao2012} is to recognize that an invertible smooth curve can be constructed with
$D(M)$
as a function of
$M$
by using a sequence of uniform iid random data. Now, since Shannon entropy
$H_{N}(X;M)$
is defined as a function of
$M$
and for equiprobable symbols, we have
\begin{equation}H_{0}(M) =\log _{2}(M)
\end{equation}
this indicates that if the unknown value of
$M$
can be estimated directly from the data, then the entropy can be determined.

A model for estimating
$M$
can be obtained by forming an appropriate, eg. polynomial model, inverting the original
equation found in (\ref{Jeqn2}), as
\begin{align}\widehat{M}(D) &  =G\left (\Theta ;D\right ) \label{Geqn} \\
 &  =\sum _{i =0}^{n_{p}}\theta _{i}D^{i}\end{align}
and appropriate values for the parameters
$\theta _{i}$
by fitting a curve to an ensemble of data. In \cite{Montalvao2012}, setting
$n_{p} =2 ,$
the values obtained were
$\theta _{0} =0.1272 ,\theta _{1} = -0.8493 ,\theta _{2} =0.6366.$
The entropy can then be estimated as
\begin{equation}\widehat{H}_{0} =\log _{2}(\widehat{M}) \label{simpleEntropy}
\end{equation}

Experimentally, this approach was shown to provide good accuracy using only a
small number of symbol coincidence distance observations \cite{Montalvao2012}. The limitation
however is the assumption of equiprobable symbol probabilities. In the next section we propose a new
algorithm which extends this method to the case of non-equiprobable symbols.

\subsection{Coincidence Counting For Non-Equiprobable Symbols
}

For natural sequences, including natural language, a mechanism to model the
non-equiprobable symbolic probabilities is to use a Zipfian law where the
probability of information events can generally be ranked into monotonically
decreasing order. For natural language, it has been shown that Zipf's law approximates the
distribution of probabilities of letter or words across a corpus of sufficient
size for the larger probabilities \cite{Piantadosi2014}. We do not rely on Zipf's law to provide a universal model of human
language or other natural sequences
(see for example, the discussions in \cite{li92random}, \cite{liw02},\cite{Corral2015}). Nevertheless,
Zipfian laws have been proven to be useful as a means of
statistically characterizing the observed behaviour of symbolic sequences of data (\cite{Montemurro2001}) and are useful in
forming a model of symbolic information transmission which is organized on the basis of sentences
made by words in interaction with each other \cite{Cancho01}. Here we adopt the
Zipf-Mandelbrot-Li law described in \cite{BackAngusWilesEstEntropy2018}, as a model for natural sequences with
non-equiprobable distribution of symbols.

In the former case, we have a model defined by
$f(n;M)$
from which a smooth invertible model
$J(n;M)$
is obtained. Thus we can obtain a model
$G\left (\Theta ;D\right )$
which enables the entropy to be estimated directly from the symbol coincidences. To
derive a model for the non-equiprobable case, one approach is to model individual
$D_{i}$
and assume some form of discrete probability related to each distance.

The method we propose is that following (\ref{feqn})-(\ref{Jeqn2}) a model
$J^{ \prime }(n;M ,r)$
is defined for each symbol, indexed by rank
$r .$
Therefore, for any given
$M ,$
each symbol of a specified rank
$r$
can be treated as being equiprobable. Thus, if the probability can be determined for
each symbol in terms of its rank, and this can be related to the overall entropy, then the same
approach can be followed as for the equiprobable case.

Consider a reformulation of (\ref{Fneqn})
where:
\begin{align}\widetilde{F}\left (n;M\right ) &  =\frac{M(M -1) \cdots (M -n +1)}{M^{n}} \nonumber  \\
 &  =\frac{M}{M} \cdot \frac{(M -1)}{M} \cdot \frac{(M -2)}{M} \cdots \frac{(M -n +1)}{M} \nonumber  \\
 &  =1 \cdot \left (1 -\frac{1}{M}\right ) \cdot \left (1 -\frac{2}{M}\right ) \cdots \left (1 -\frac{n -1}{M}\right ) \nonumber  \\
 &  =1 \cdot \left (1 -P_{2}\right ) \cdot \left (1 -P_{3}\right ) \cdots \left (1 -P_{n -1}\right ) \label{FPfactor}\end{align}
using the identity
$(M -n +1)/M =1 -(n -1)/M$
and
$P_{h}$
is the probability of independently drawing\protect\footnote{ If this was cast in the classic case of drawing colored objects from a
bag, it would be with replacement.
}
$h -1$
identical symbols from a set of
$M$
in
$h -1$
draws. In the case of equiprobable symbols, we have
\begin{equation}\widetilde{P}_{h}(M) =1 -\frac{h -1}{M}
\end{equation}
Now, for a natural sequence where the probability of occurrence of a given word can be
defined in terms of rank, the Zipf-Mandelbrot-Li law provides an expression for the probability to
be used in (\ref{FPfactor}) where (\cite{BackAngusWilesEstEntropy2018},\cite{Montemurro2001},\cite{Mandelbrot83}):
\begin{equation}P(r;M) =\frac{\gamma _{}^{ \prime }}{\left (r +\beta _{}\right )^{\alpha _{}}} \label{prank}
\end{equation}
and for iid samples, the constants can be computed as (\cite{li92random}):
\begin{equation}\alpha  =\frac{\log _{2}(M +1)}{\log _{2}(M)} ,\beta  =\frac{M}{M +1} ,\gamma _{M} =\frac{M^{\alpha  -1}}{\left (M -1\right )^{\alpha }}
\end{equation}
and
\begin{equation}\gamma _{}^{ \prime } =\frac{\gamma _{}}{\kappa _{}}
\end{equation}
where
\begin{equation}\sum \limits _{i =1}^{M}p(i) =1 ,\quad \sum \limits _{i =1}^{M}\frac{\gamma _{}}{(r +\beta _{})^{\alpha _{}}} =\kappa _{}
\end{equation}
This approach provides an equiprobable representation of the symbols by considering a
different model for each symbol rank, according to the rank. But moreover, once a model is found for
one rank, then the whole model can be identified. Hence, adopting a probabilistic model according to
the symbolic rank we define
\begin{equation}F\left (n;r ,M\right ) =1 -\prod \limits _{h =1}^{n}\left (1 -P_{h}(r ,M)\right )
\end{equation}
where
\begin{equation}P_{h}(r ,M) =\frac{h\gamma ^{ \prime }}{(r +\beta )^{\alpha }}
\end{equation}
Therefore, the same approach as before can be adopted by defining
\begin{equation}f(n;r ,M) =F(n;r ,M) -F(n -1;r ,M) \label{frank}
\end{equation}

Hence, we now have
$E_{r}[n] =J^{ \prime }(n;r ,M)$
and
\begin{equation}D_{r}(M) =\sum \limits _{n =0}^{M}nf(n;r ,M) \label{Jprimeeqn}
\end{equation}
Using a similar approach to the previous equiprobable case, a per symbolic rank model
for estimating
$M$
can be obtained by prescribing\protect\footnote{Note that although it is technically feasible to derive the exact model
$J^{ \prime }(n;r ,M)$
in terms of (\ref{prank})-(\ref{frank}), it is not necessary to do so in practice as is evident by the
curve fitting approach proposed in \cite{Montalvao2012} and adopted here.
}
$J^{ \prime }(n;r ,M)$
in (\ref{Jprimeeqn}), and then  inverting this
to become
\begin{equation}\widehat{M_{r}}(D) =G_{r}\left (\Theta ;M ,D_{r}\right )
\end{equation}
Now, unlike the model proposed initially in \cite{Montalvao2012}, natural sequence data
consists of a non-equiprobable set of symbols
and so we cannot simply use (\ref{simpleEntropy}) to estimate entropy in a single step as before. However, given
an estimate
$\widehat{M_{r}}(D) ,$
from the observed inter-symbol distance, it now becomes possible to apply this
parameter to the Zipf-Mandelbrot-Li set of equations in addition to our rank-based probability
model of symbol drawings, and obtain an overall estimate for the entire set of symbolic
probabilities. While this can be achieved using, for example,
$D_{1}$
, clearly it is possible to form an estimate which uses
$D_{i}$
for
$i =1..n$
according to any desired criteria such as least squares or any other norm. Having then
estimated
$\widehat{P}_{h}(r ,M) ,$
the entropy can then be easily estimated as
\begin{equation}\widehat{H}_{1}(r ,X) = -\sum \limits _{h =1}^{\widehat{M}}\widehat{P}_{h}(r ,M)\log _{2}\left (\widehat{P}_{h}(r ,M)\right )
\end{equation}
which defines the rank
$r$
Shannon entropy estimate.  In the next section, we demonstrate the performance of the
model in various simulations.

\section{Example Results
}

\subsection{Synthetic Entropy Model of English Text
}
 In this example, a set of data is simulated using the Zipf-Mandelbrot-Li model
with 27 symbols corresponding to the 26 letters and a space. The rank-based entropy estimation
algorithm described in the previous section is used to estimate the model by counting the
coincidences of the symbols. In the first instance, we simply compute the average symbol distance
$D_{1}$
and then apply this to the inverted model. Note that a different model applies to each
rank as shown in Fig. 1.
\begin{figure}\centering \includegraphics[bb=38 214 575 584, width=8cm,]{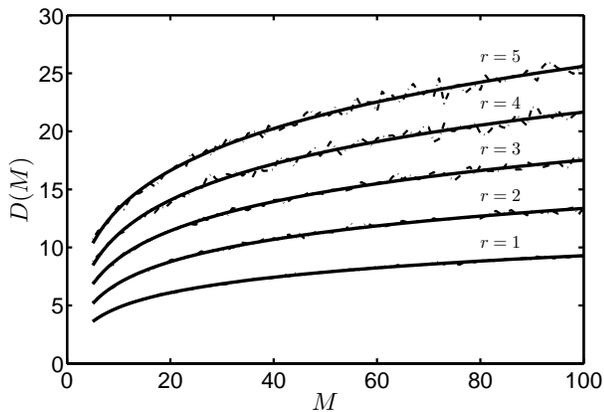}\caption{Rank-based entropy models for
$D(M) =J^{ \prime }(n ,r ,M)$
. Note that the symbol distances are measured according to their rank.
}\end{figure}
The rank-based entropy models for
$D(M) =J^{ \prime }(n ,r ,M)$
are inverted and the models are shown in Fig. 2. Here, a power based model
is used,
\begin{equation}\widehat{M_{r}}(D) =aD_{r}^{b} +c
\end{equation}
where
$a =\ensuremath{\operatorname*{0.0075}} ,b =\ensuremath{\operatorname*{4.2345}} ,c =\ensuremath{\operatorname*{4.1385}} .$
In the synthetic simulation results, using only 25 symbol coincidences, where the true
entropy is
$H_{a}(27) =4.261$
by application of the rank-based entropy model described in the previous section, we
obtain the estimated entropy of
$H_{e}(27) =4.266$
indicating the efficacy of the method.

\subsection{Entropy of English Text: Tom Sawyer
}
 In this example, the classic English language text Tom Sawyer was used to test
the algorithm. In this case, the rank 1 model was again used, where the highest ranked symbol
corresponds to the space character. Commencing at Chapter 2 of the text, the intersymbol distance
was estimated as
$D_{1}(50) =6.03$
which leads to an estimated entropy of
$H_{e}(27) =4.3$
which is in close agreement to the actual entropy of the text where
$H_{a}(27) =4.4$.
\begin{figure}\centering \includegraphics[bb= 34 205 569 589, width=8cm,]{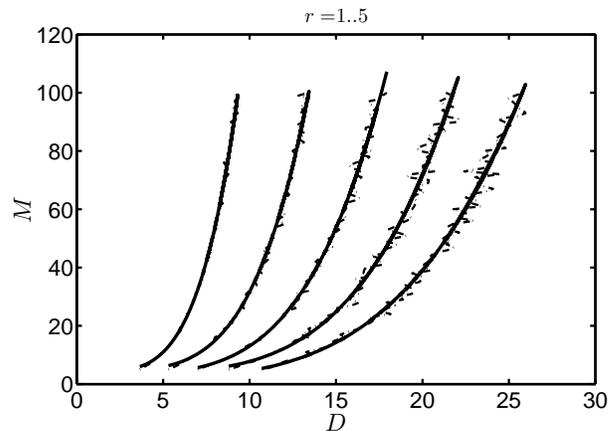}\caption{Inverse rank-based entropy models for
$\widehat{M_{r}}(D) =G_{r}\left (\Theta ;M ,D_{r}\right ) .$
Each model is derived from the initial rank based model which describes the
symbolic distance
$D_{r}$
as a function of
$M.$
}\end{figure}
Moreover, the result was obtained by using less than 300 characters
or 50 words which is quite remarkable.

\section{Conclusion
}
Shannon entropy is a well known method of measuring the information
content in a sequence of probabilistic symbolic events. In this paper, we have proposed a fast
algorithm for estimating Shannon Entropy for natural sequences.
Using a modified Zipf-Mandelbrot-Li law and a coincidence counting method, we have demonstrated a
method which gives extremely fast performance in comparison to other techniques and
yet is simple to implement. Examples have been given which show the efficacy of the proposed
methodology. It would be of interest to apply this method to various real world applications to
compare the theoretical results against experimentally obtained results. In terms of information
theoretic analytical tools, it may be of interest to consider just how few samples may be required
in order to obtain useful results. In order to make the most use of available data, future work
could consider optimal strategies for deriving accurate models from multiple symbol ranks; this
could be expected to yield fruitful results especially when there is some `noise' in the data, eg
some symbols are missing. Another area of interest in future work will be to analyze the bias of the
model as considered in \cite{Schurmann2004}.

\subsection*{Acknowledgments
}
The authors would like to acknowledge partial support from the Australian Research
Council Centre of Excellence for the Dynamics of Language and helpful discussions with Dr Yvonne Yu
and Dr Paul Vrbik.
\bibliographystyle{apsrev4-1}
%%\bibliography{tap1}

\end{document}